\documentclass[preprint,showpacs,preprintnumbers,amsmath,amssymb]{revtex4}

\usepackage[dvipdfmx]{graphicx}
\usepackage{dcolumn}
\usepackage{bm}
\usepackage{color}

\begin{document}

\newif\ifplot
\plottrue
\newcommand{\RR}[1]{[#1]}
\newcommand{\intsum}{\sum \kern -15pt \int}
\newfont{\Yfont}{cmti10 scaled 2074}
\newcommand{\Y}{\hbox{{\Yfont y}\phantom.}}
\def\O{{\cal O}}
\newcommand{\bra}[1]{\left< #1 \right| }
\newcommand{\braa}[1]{\left. \left< #1 \right| \right| }
\def\Bra#1#2{{\mbox{\vphantom{$\left< #2 \right|$}}}_{#1}
\kern -2.5pt \left< #2 \right| }
\def\Braa#1#2{{\mbox{\vphantom{$\left< #2 \right|$}}}_{#1}
\kern -2.5pt \left. \left< #2 \right| \right| }
\newcommand{\ket}[1]{\left| #1 \right> }
\newcommand{\kett}[1]{\left| \left| #1 \right> \right.}
\newcommand{\scal}[2]{\left< #1 \left| \mbox{\vphantom{$\left< #1 #2 \right|$}}
\right. #2 \right> }
\def\Scal#1#2#3{{\mbox{\vphantom{$\left<#2#3\right|$}}}_{#1}
{\left< #2 \left| \mbox{\vphantom{$\left<#2#3\right|$}}
\right. #3 \right> }}

\title{Proton-$^3\rm He$ elastic scattering at intermediate energies 
}

\author{A.\ Watanabe$^{1}$}
\email{watanabe@lambda.phys.tohoku.ac.jp}
\author{S.\ Nakai$^{1,2}$}
\email{nakai@lambda.phys.tohoku.ac.jp}
\author{Y.\ Wada$^{1}$}
\author{K.\ Sekiguchi$^{1}$}
\email{kimiko@lambda.phys.tohoku.ac.jp}
\author{A. Deltuva$^{3}$}
\author{T.\ Akieda$^{1}$}
\author{D.\ Etoh$^{1}$}
\author{M.\ Inoue$^{1}$}
\author{Y.\ Inoue$^{1}$}
\author{K.\ Kawahara$^{1}$}
\author{H.\ Kon$^{1}$}
\author{K.\ Miki$^{1}$}
\author{T.\ Mukai$^{1}$}
\author{D.\ Sakai$^{1}$}
\author{S.\ Shibuya$^{1}$}
\author{Y.\ Shiokawa$^{1}$}
\author{T.\ Taguchi$^{1}$}
\author{H.\ Umetsu$^{1}$}
\author{Y.\ Utsuki$^{1}$}
\author{M.\ Watanabe$^{1}$}
\author{S. Goto$^{4}$}
\author{K. Hatanaka$^{5}$}
\author{Y. Hirai$^{4}$}

\author{T. Ino$^{6}$}
\author{D.\ Inomoto$^{4}$}
\author{A. Inoue$^{5}$}
\author{S.\ Ishikawa$^{7}$}
\author{M. Itoh$^{8}$}

\author{H. Kanda$^{5}$}
\author{H.\ Kasahara$^{4}$}
\author{N. Kobayashi$^{5}$}
\author{Y.\ Maeda$^{9}$}
\author{S.\ Mitsumoto$^{4}$}
\author{S. Nakamura$^{5}$}
\author{K.\ Nonaka$^{9}$}
\author{H. J. Ong$^{5,10}$}
\author{H.\ Oshiro$^{4}$}
\author{Y.\ Otake$^{11}$}
\author{H.\ Sakai$^{12}$}
\author{A.\ Taketani$^{11}$}
\author{A. Tamii$^{5}$}
\author{D.\ T.\ Tran$^{5}$}
\author{T.\ Wakasa$^{4}$}
\author{Y.\ Wakabayashi$^{11}$}
\author{T.\ Wakui$^{13}$}

\affiliation{$^{1}$
Department of Physics, Tohoku University, Sendai 980-8578, Japan}
\affiliation{$^{2}$
Graduate Program on Physics for the Universe (GP-PU), Tohoku University, Sendai 980-8578, Japan
}%
\affiliation{$^{3}$
Institute of Theoretical Physics and Astronomy, Vilnius University, 
Saul\.{e}tekio al. 3, LT-10257 Vilnius, Lithuania
}%
\affiliation{$^{4}$
Department of Physics, Kyushu University, Fukuoka 819-0395, Japan
}
\affiliation{$^{5}$
Research Center for Nuclear Physics, Osaka University,
Ibaraki 567-0047, Japan
}
\affiliation{$^{6}$
High Energy Accelerator Research Organization (KEK), Tsukuba 305-0801, Japan
}%
\affiliation{$^{7}$
Science Research Center, 
Hosei University, Tokyo 102-8160, Japan}%
\affiliation{$^{8}$
Cyclotron and Radioisotope Center, Tohoku University, Sendai 980-8578, Japan
}%
\affiliation{$^{9}$
Faculty of Engineering, University of Miyazaki, Miyazaki 889-2192, Japan
}
\affiliation{$^{10}$
Institute of Modern Physics (IMP), Chinese Academy of Sciences, Lanzhou 730000, China
}
\affiliation{$^{11}$
RIKEN Center for Advanced Photonics, Wako 351-0198, Japan
}%
\affiliation{$^{12}$
RIKEN Nishina Center, Wako 351-0198, Japan
}%
\affiliation{$^{13}$
National Institute of Radiological Science, Chiba 263-8555, Japan
}%

\date{\today}

\begin{abstract}
\indent 
We present a precise measurement of the cross section, proton and $\rm ^3He$ analyzing powers, and spin correlation coefficient $C_{y,y}$ for $p$-$\rm ^3He$ elastic scattering 
near 65 MeV,  and a comparison 
with rigorous four-nucleon scattering calculations 
based on realistic nuclear potentials and a model with $\Delta$-isobar excitation. 
Clear discrepancies are seen in some of the measured observables in the regime around the cross section minimum. 
Theoretical predictions using scaling relations between the calculated cross section and 
the $\rm ^3 He$ binding energy are not successful in reproducing the data. 
Large sensitivity to the $NN$ potentials 
and rather small $\Delta$-isobar effects in the calculated cross section 
are noticed as different features from those in the deuteron-proton elastic scattering.  
The results obtained above indicate that $p$-$\rm ^3He$ scattering at intermediate energies 
is an excellent tool to explore nuclear interactions not accessible by three-nucleon scattering.

\end{abstract}

\maketitle

\section{Introduction}


One of the open questions in nuclear physics nowadays is a 
complete knowledge of the interactions acting among nucleons.
Modern nucleon-nucleon ($NN$) potentials reproducing 
the $NN$ observables up to 350~MeV 
with very high precision 
do not describe that well  various nuclear phenomena,
e.g., some few-nucleon scattering observables, nuclear binding energies,
and nuclear matter properties~\cite{glo96,Carlson.Rev.Mod.Phys.87.1067}.
The three-nucleon forces (3$N$Fs),
arising naturally in the standard 
meson-exchange picture~\cite{fuj57}
as well as in  the chiral effective field theory 
($\chi$EFT)~\cite{kolk1994,epel2009},
have been suggested as possible candidates
to improve the situation. 
Few-nucleon reactions offer good opportunities 
to investigate the nature of the nuclear interactions
since rigorous numerical 
calculations with two-nucleon ($2N$) and $3N$ forces
and high-precision experiments are feasible.

Nucleon-deuteron $(Nd)$ elastic scattering at energies 
above $\approx 60~\rm MeV/nucleon$ has been considered as a solid basis 
to explore the nuclear interactions focusing on the 3$N$Fs.
Sizable discrepancies between the data and  rigorous numerical 
calculations with realistic $NN$ potentials were
found in the cross section minimum \cite{nsakamot96}.
They were successfully explained by inclusion of the two-$\pi$ exchange $3N$F models
that reproduce the binding energies of $^3\rm H$ and $^3\rm He$~\cite{wit98,sakai2000,wit2001},
or substantially reduced by calculations in an extended Hilbert space 
which allow the explicit excitation of a nucleon to a $\Delta$ isobar, 
yielding an effective $3N$F~\cite{nemoto98,CDBD}.
It has been recently reported that the deuteron-proton ($d$-$p$) elastic cross section 
at 70 MeV/nucleon~\cite{sekiguchi2002}  constrains
low-energy constants of 3$N$Fs in $\chi$EFT~\cite{Epelbaum2019}.  

The four-nucleon ($4N$) system has also become a test field
 for modern nuclear forces. 
It is the simplest system of investigating
the nuclear interactions in $3N$ subsystems with the total isospin $T = 3/2$,
whose importance
is suggested in asymmetric nuclear systems, e.g., neutron-rich nuclei~\cite{piep2001} 
and pure neutron matter~\cite{n_matter1,n_matter2}. 
In recent years remarkable theoretical studies in solving the $4N$ 
scattering problem with realistic Hamiltonians have been reported
\cite{Deltuva_PRC76,lazauskas2009,viviani_PRL111} 
even above the $4N$ breakup threshold~\cite{deltuva_prc87,fonseca_FBS2017}, 
opening new possibilities for nuclear force study 
in the $4N$ system at intermediate energies.

Thanks to developments in technology for 
high quality polarized and unpolarized proton beams
together with sophisticated techniques for the polarized 
$^3\rm He$ target system, 
$p$-$^3\rm He$ scattering has an experimental advantage
that allows high precision measurements of the cross section
and a variety of spin observables. 
Indeed, at proton energy below $50~\rm MeV$ rich data sets 
for $p$-$^3\rm He$ elastic scattering are available, covering the cross section
\cite{p3He9p75,p3He1p20,p3He6p5,p3He8.5,McDonald,CCKim,p3He2.38,Harbison,Hutson,p3HeMurdoch,p3HeViviani,Fisher},  
proton analyzing power~\cite{McDonald,Harbison,Baker,Jarmie,Birchall,Alley,p3HeViviani,Fisher}, 
$\rm ^3He$ analyzing power~\cite{Baker,Alley,Muller,McCamis,Daniels}, 
and spin-correlation coefficients~\cite{Baker,Alley,Daniels}. 
However,  the existing data basis is rather poor at higher energies, mostly limited to
the cross section~\cite{Votta,Goldstein,Wesick,Langevin-Joliot}
and the proton analyzing power~\cite{Wesick}. 
Few data exist for the $^3\rm He$ analyzing power and the spin correlation coefficients~\cite{Shimizu}.

As for the theoretical descriptions of $p$-$^3\rm He$ elastic scattering, 
calculations taking into account the $2N$ and $3N$ forces 
are reported 
in the framework of the Kohn variational approach
up to 5.54 MeV (below $3N$ breakup threshold) ~\cite{viviani_PRL111}. 
At these lower energies, the data are well explained by the calculations 
with the $2N$ interactions.
The exception is the proton analyzing power 
$A_y$, for which the so-called $A_y$ puzzle exists as seen 
in the $Nd$ elastic scattering.
At higher energies, calculations in the framework of 
the Alt-Grassberger-Sandhas (AGS) equation are presented 
using various realistic $NN$ potentials up to 35 MeV~\cite{deltuva_prc87,fonseca_FBS2017}. 
Although calculations including $3N$Fs have not been done  
for energies above the breakup threshold 
so far, these works performed coupled-channel
calculations with the $\Delta$-isobar degree of freedom 
as an alternative source  of $3N$ and $4N$ forces. 
At incident energies above 20 MeV, the calculations with realistic $NN$ potentials 
underpredict the cross section data in the minimum, 
as observed also in the $Nd$ elastic scattering but at higher center-of-mass energy. 
The $\Delta$-isobar effects slightly improve the agreement with the data for the cross section.
In line with this feature it would be interesting to see how the theoretical calculations 
based on realistic nuclear potentials explain the data for $p$-$^3\rm He$ 
elastic scattering at intermediate energies. 
 
In this paper we present the first precise data set 
for $p$-$^3\rm He$ elastic scattering at intermediate energies,
the cross section $d\sigma/d\Omega$ and 
the proton analyzing power $A_y$ at 65 MeV, and
the $^3\rm He$ analyzing power $A_{0y}$ at 70 MeV
spanning a wide angular range.
In addition, we present 
the spin correlation coefficient $C_{y,y}$ at 65 MeV at 
the angles of $46.6^\circ$, $89.0^\circ$, $133.2^\circ$
in the center-of-mass system.
The data are compared with rigorous numerical 
calculations for $4N$-scattering
based on various realistic $NN$ potentials
as well as  with the $\Delta$-isobar excitation,
in order to explore possibilities of $p$-$^3\rm He$ elastic scattering
as a tool to study nuclear interactions.
However, the Coulomb force is omitted this time, thus 
the theoretical predictions of the present work in fact refer to the mirror
reaction $n$-$^3\rm H$.

In Sec.~II, we describe the experimental procedure and the data analysis. 
Section III presents a comparison between the experimental data 
and the theoretical predictions, and discussion follows in Sec.~IV.  
Finally, we summarize and conclude in Sec.~V.

\section{Experimental procedure and data analysis}

\subsection{Measurement of the cross section and the proton analyzing power $A_y$ at 65~MeV}

\begin{figure}[h]
	\centering
	\includegraphics[scale=0.6]{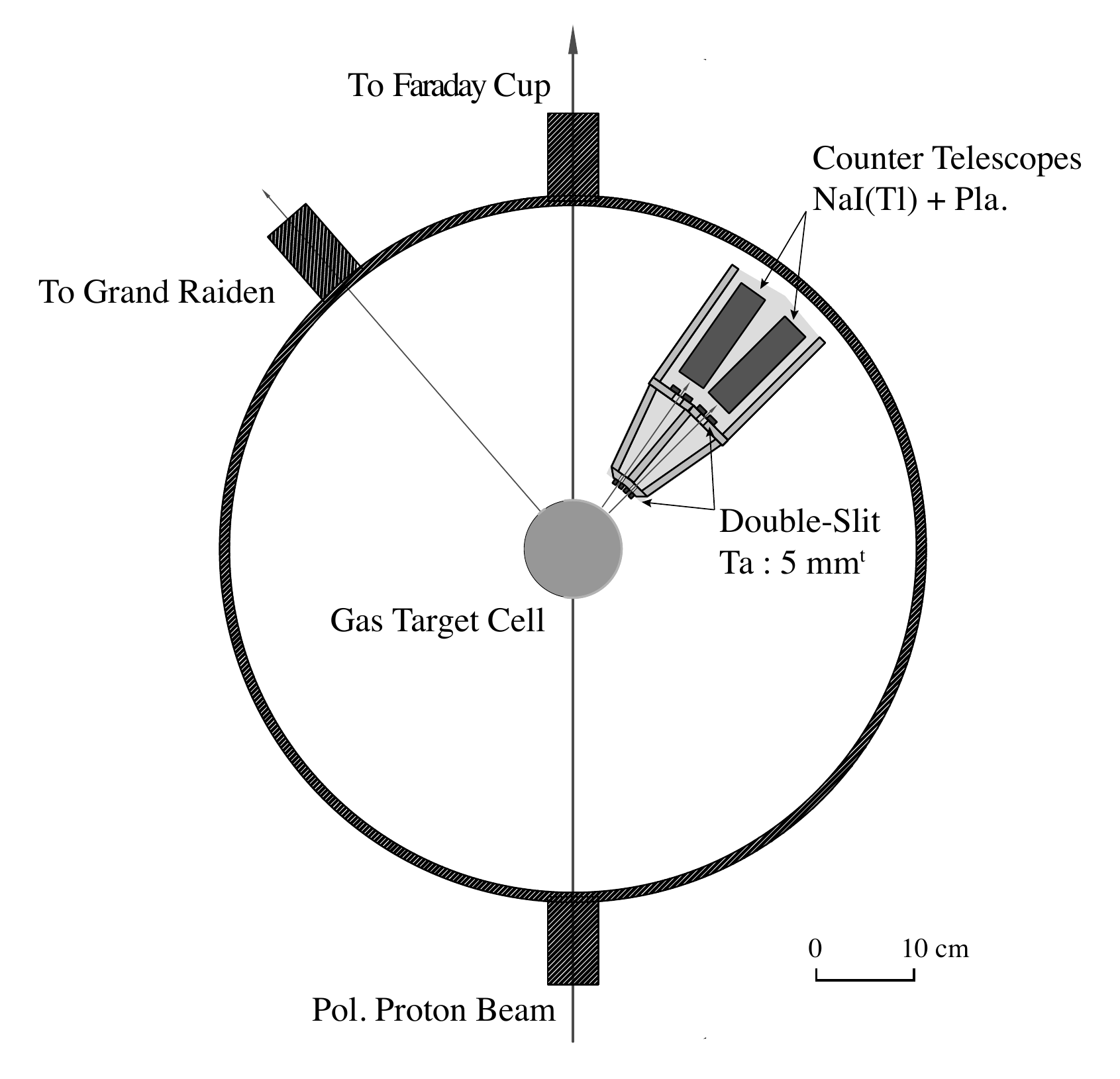}
	\caption{
	Schematic layout of the experimental setup for the measurement of the cross section and 
	the proton analyzing power $A_{y}$. }
    \label{fig:xs_setup}
\end{figure}

\begin{figure}[h]
	\centering
	\includegraphics[scale=0.4]{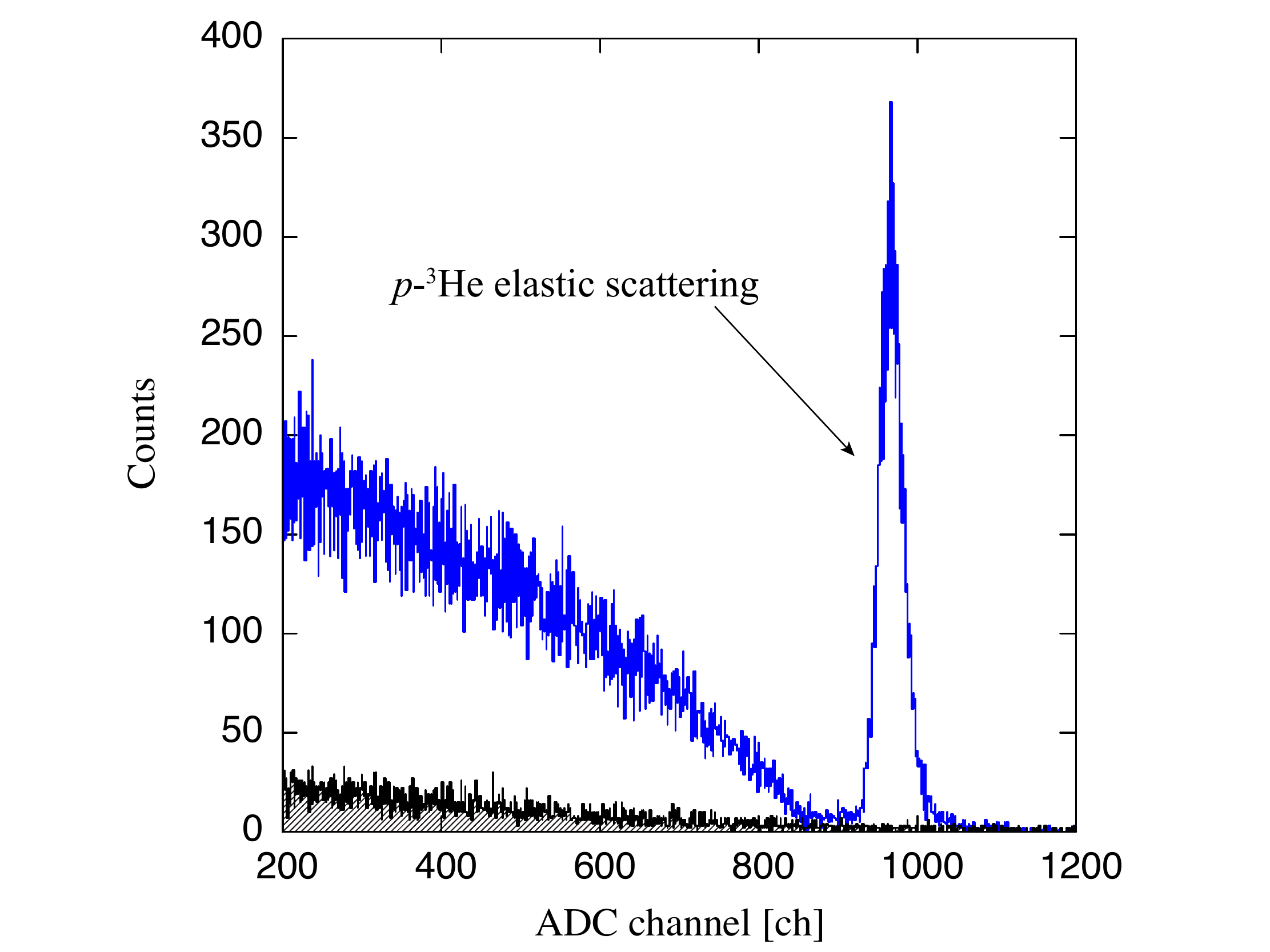}
	\caption{Light output spectrum of scattered protons
	obtained by the NaI(Tl) scintillator at $\theta_{\rm lab.}$ = 75$^{\circ}$.
	The hatched region indicates events obtained with the empty target cell.}
    \label{fig:xs_spectrum}
\end{figure}

The measurement of the cross section $d\sigma/d\Omega$ 
and the proton analyzing power $A_y$
was performed with a 65 MeV polarized 
proton beam in the West Experimental Hall 
at the Research Center for Nuclear Physics (RCNP), Osaka University.
The measured angles 
were $\theta_{\rm lab.} = 20.0^\circ$--$165.0^\circ$ in the laboratory system
which corresponds to 
$\theta_{\rm c.m.} = 26.9^\circ$--$170.1^\circ$ in the center-of-mass system.
Figure~\ref{fig:xs_setup} shows 
the schematic layout of the experimental setup around the $^3\rm He$ target. 
The polarized proton beam provided by an atomic beam type
polarized ion source~\cite{PIS_RCNP} 
was accelerated up to 65 MeV by the AVF cyclotron.
The beam was transported~\cite{WS_NIM} 
to the $^3\rm He$ gas target 
at the center of the scattering chamber 
of the magnetic spectrometer Grand Raiden~\cite{GR_NIM}
that was used to monitor the luminosity 
by observing the $p$-$^3\rm He$ elastic scattering events at a laboratory angle of $41^\circ$.
The beam was stopped in a Faraday cup located 
downstream of the scattering chamber.
The beam intensity was 20--100 nA.
The beam polarization was 
monitored with a beamline polarimeter of the West Experimental Hall
by using $p$-$^{12}\rm C$ elastic scattering~\cite{ieiri_NIM}.
During the measurement,  the beam polarization was typically 53\%. 
The $^3\rm He$ gas target was contained in the cell of
a cylinder of 99 mm diameter with a 50-$\mu \rm m$-thick
Al window 
with a pressure of 1 atm at room temperature. 
The absolute gas density was determined 
with uncertainty of less than 0.8\%
by continuously monitoring the pressure 
as well as the temperature during the measurement.
%
Scattered protons from the $^3\rm He$ target were
detected with two sets of counter telescopes 
which were placed in the scattering chamber
and positioned 20~cm away from the center of the target cell.
Each counter telescope consisted of 
a NaI(Tl) scintillator 
with dimensions 50 mm (thickness) $\times$ 31 mm (width) $\times$ 31 mm (height)
and a 0.5-mm-thick plastic scintillator.
A double-slit system 
was used to define the target volume and the solid angle.
Each slit was made of 5-mm-thick Ta.
The effective target thickness and the solid angle were 
calculated by Monte Carlo simulations.
%
%
%
%
%
%
%
%
Elastically scattered protons from the $^3\rm He$ target
were identified using the correlation between 
energy loss of a plastic scintillator
and the remaining energy deposited in a NaI(Tl) scintillator. 
Figure~\ref{fig:xs_spectrum} shows the light output spectrum 
for the scattered protons obtained by a NaI(Tl) scintillator 
at the laboratory angle of $70^\circ (\theta_{\rm c.m.}= 89.0^\circ $). 
%
%
The peak corresponding to protons elastically scattered from $^3\rm He$ 
was well separated from the inelastic scattering events.
%
The background events
obtained with the empty target cell showed 
a nearly flat distribution.
After subtracting the background contributions
the yields of the $p$-$^3\rm He$ elastic scattering 
were extracted by fitting with skewed Gaussians.
%
The effects of nuclear reactions in the 
NaI(Tl) scintillators 
were interpolated from the published results~\cite{NaI_eff_1, NaI_eff_2}.
%
%
The absolute values of the cross section 
for the $p$-$^3\rm He$ elastic scattering 
were deduced by normalizing the data to
the $p$-$p$ scattering cross section as given
by the phase-shift analysis program SAID~\cite{SAID}.
The normalization factors were obtained 
from the measurement for the $p$-$p$ scattering 
with hydrogen gas by using the same detection system 
for the $p$-$^3\rm He$ scattering. 
%
The statistical error of $d\sigma/d\Omega$
for the $p$-$^3\rm He$ elastic scattering
is better than $\pm 2 \%$.
The systematic uncertainty, which is the quadratic sum
of 
the uncertainty in the normalization factor, 
the uncertainty in the background contamination,
the fluctuation  in the luminosity,
and the uncertainty in the beam polarization
is estimated to be $3 \%$.
For $A_y$, the statistical error is 0.02 
or less and the systematic uncertainty is estimated to be 0.02.

\subsection{Measurement of the $^3\rm He$ analyzing power $A_{0y}$ at 70 MeV}

\begin{figure}[htbp]
	\includegraphics[scale=0.3]{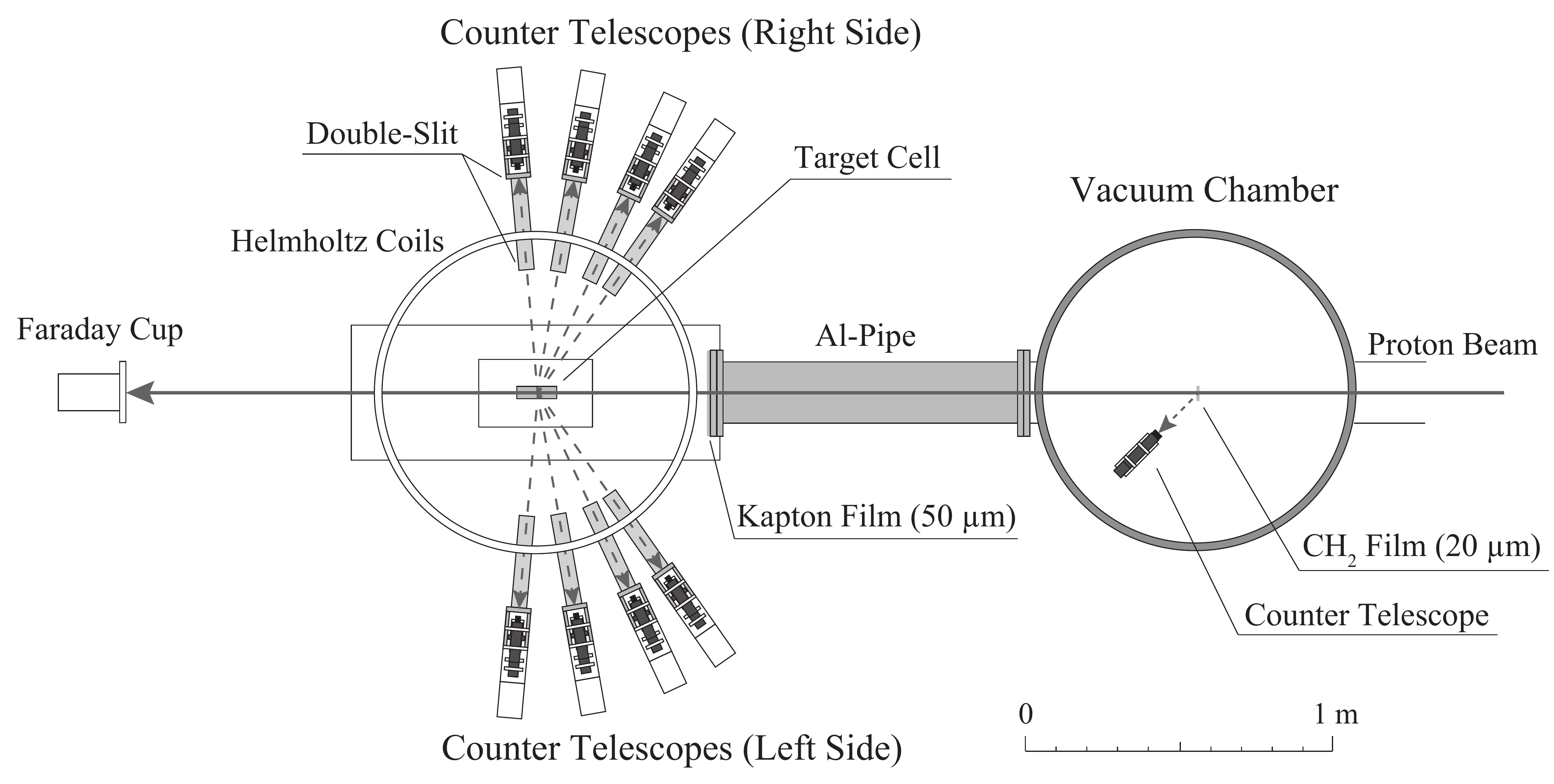}
	\caption{Schematic layout of 
	the experimental setup for the measurement 
	of the $^3\rm He$ analyzing power $A_{0y}$.}
	\label{fig:A0y_setup}
\end{figure}

The measurement of the $^3\rm He$ analyzing power $A_{0y}$ was performed 
with a 70 MeV proton beam in conjunction with 
the polarized $\rm^3He$ target  
at the Cyclotron Radioisotope Center (CYRIC), Tohoku University.
The experiment consisted of several separate measurements 
by using two different polarized $^3\rm He$ target cells.
The experimental setup around the target is shown in Fig.~\ref{fig:A0y_setup}.
A proton beam with an intensity of 5--10 nA bombarded 
the polarized $^3\rm He$ target and it was stopped in a Faraday cup. 
%
%
%
Relative beam intensity was monitored by 
a beam monitoring system installed in the vacuum chamber,
by which scattered protons from a polyethylene film 
with a thickness of 20 $\mu$m were detected.
The polarized $^3\rm He$ target 
and the detector system for the $p$-$^3\rm He$ scattering were 
operated in atmosphere.
The vacuum was separated by a Kapton film 
with thickness of 50~$\mu$m which was attached to 
an aluminum made beam pipe connected to the vacuum chamber.
%
%
Scattered protons from the $^3\rm He$ target
were detected using sets of counter telescopes
which were positioned 73~cm away from the center of the target cell
symmetrically on each side of the beam axis
at laboratory angles of $35.0^\circ$--$125.0^\circ$
($\theta_{\rm c.m.}=46.6^\circ$--$141.4^\circ$). 
Each counter telescope consisted of a NaI(Tl) scintillator 
and a plastic scintillator. 
The NaI(Tl) scintillator was the same type
as was used for the cross section measurement.
%
For the plastic scintillators,
different thicknesses, namely 0.2, 0.5, and 1.0 mm, 
were used depending on the measured angles.
A double-slit collimator,
which was made of 20-mm-thick Al for the front part and
15-mm-thick brass for the rear part,
was used to define the target volume and the solid angle
for each counter telescope.
%
%
%
The method to polarize a $^3\rm He$ nucleus was 
based on the principle of spin-exchange optical pumping (SEOP)~\cite{PRL.5.373, PRL.91.123003}. 
%
%
A target cell was one-piece GE180 glassware which consisted of a
pumping chamber and a target chamber, connected by a thin transfer tube. 
This design prevented the depolarization of alkali-metal atoms due to the incident beam~\cite{COULTER198929}.
In addition to this, undesirable energy loss of the scattered protons passing through a material, 
which is used to heat the target cell for SEOP, can be avoided.
The target cell contained $^3\rm He$ gas with a pressure of 3 atm at room temperature,
a small amount of $\rm N_2$ gas (0.1~atm), and a mixture of Rb and K alkali metals.
The pumping chamber was 
heated up to about $\rm 500~K$ to provide sufficient high alkali-metal vapor 
density and maintain the $^3\rm He $  polarization. 
Circularly polarized laser light at 794.7~nm polarized Rb atoms in the 
pumping chamber. 
$^3$He nuclei were polarized through spin exchange interactions in the pumping chamber 
and then diffused into the target chamber.
A 12 G magnetic field, provided by a pair of Helmholtz coils 100 cm in diameter,
defined the direction of the $^3\rm He$ nuclear polarization. 
The target chamber of the target cell 
had a diameter of 4 cm and was 15 cm long along the beam path.
The entrance and exit windows were made as thin as 0.4 mm,
and the thickness of the side surfaces where scattered protons passed
was about 1 mm. 
During the measurement
the target polarization was measured by 
the adiabatic fast passage nuclear magnetic resonance (NMR) method, which was calibrated 
using the electron paramagnetic resonance technique~\cite{PRA.58.3004}. 
Additionally, the absolute values of the target polarization
were measured by the thermal neutron transmission 
using the RIKEN Accelerator-Driven Compact Neutron Source (RANS)
~\cite{RANS}.
The typical target polarization was 40\% with an uncertainty of 2\%. 
A more detailed description of the target system is 
found in Ref.~\cite{Atomu_PhD}.
The statistical error of the $A_{0y}$
is 0.02 or less. The systematic uncertainty 
which mainly came from the uncertainty of the target polarization
is 0.02 or less.
The data taken independently with the different target setups
at the same angles are consistent each other within the estimated uncertainty.
%

\subsection{Measurement of the spin correlation coefficient $C_{y,y}$ at 65 MeV}

The measurement was extended to the spin correlation coefficient $C_{y,y}$.
%
%
%
%
The experiment was performed in the East Experimental Hall at RCNP.
A 65 MeV polarized proton beam with an intensity of 10 nA
bombarded the polarized $^3\rm He$ target installed at the ENN beam line~\cite{Atomu_PhD}. 
The beam polarizations monitored 
by using the proton-deuteron elastic scattering~\cite{NPA382_shimizu}
were 50\% for spin-up and 20\% for spin-down, respectively.
The same polarized $^3\rm He$ target system and 
the same detection system as those for 
the $^3\rm He$ analyzing power measurement at CYRIC were applied. 
The target polarization was  40\% during the measurement.
%
%
%
The measured angles were  
$\theta_{\rm lab.} = 35.0^\circ$, $70.0^\circ$, and $115.0^\circ$
($\theta_{\rm c.m.}=46.6 ^\circ$, $89.0 ^\circ$, and $133.2 ^\circ$). 
The statistical error of the $C_{y,y}$ varies 0.03--0.06 depending on the measured angles,
and the systematic error does not exceed the statistical one.

\section{Comparison of Data with Theoretical Calculations}

\begin{figure}[htbp] 
\includegraphics[scale=0.5]{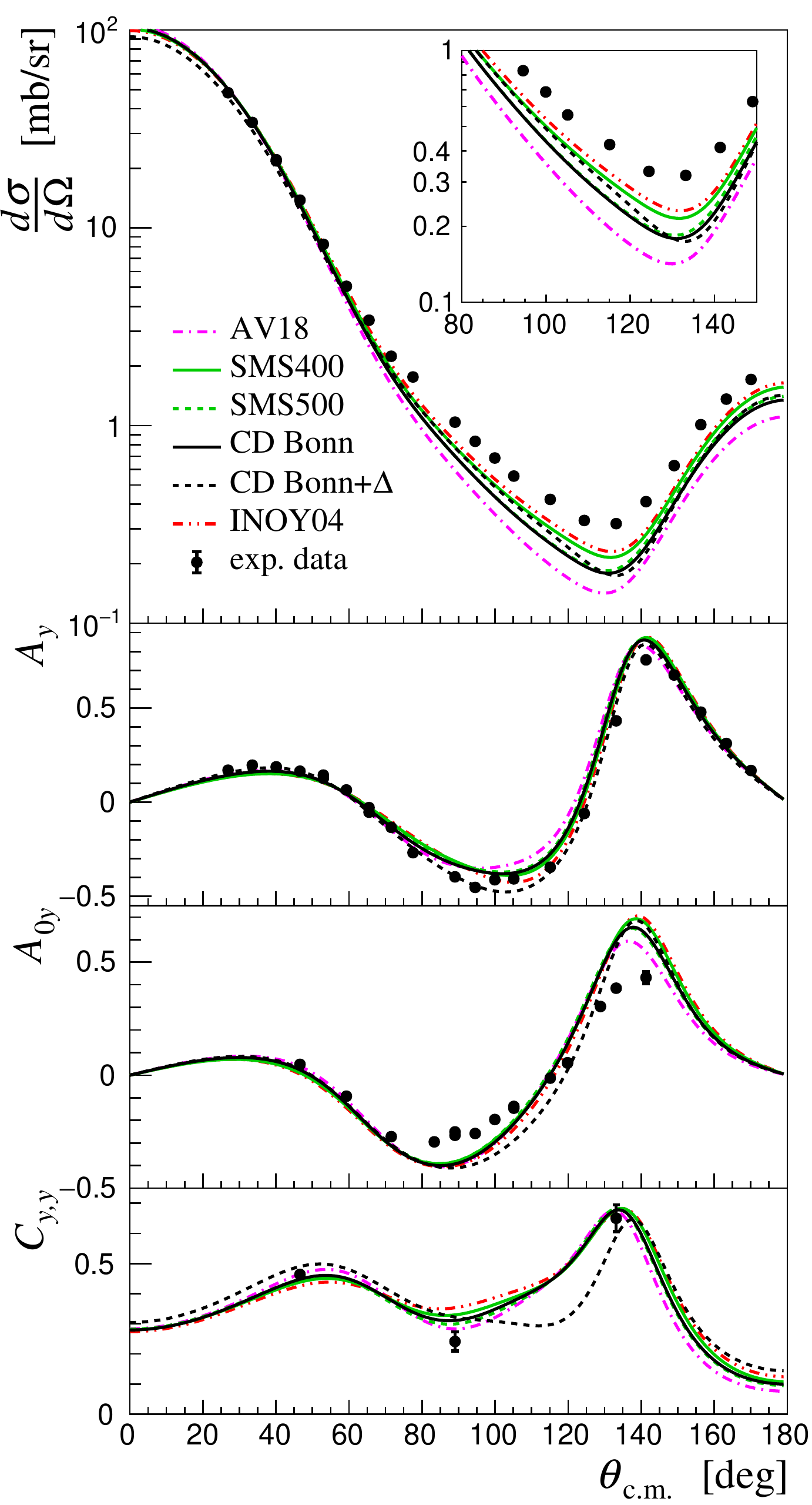}
\caption{(Color online)
Angular distributions of $d\sigma/d\Omega$ 
as well as $A_{y}$ at 65 MeV,
$A_{0y}$ at 70 MeV, and $C_{y,y}$ at 65 MeV 
for $p$-$^3\rm He$ elastic scattering.
Experimental data (solid circles) are compared 
with the calculations from the solutions of exact AGS equations.
Only statistical errors are indicated.
Calculations based on the $NN$ potentials are shown with 
magenta dash-dotted (AV18), black solid (CD Bonn), 
red dot-dot-dashed (INOY04), 
green solid (SMS400), and green dashed (SMS500) lines.
Black dashed lines are calculations based on the CD Bonn+$\Delta$ potential.
%
%
%
\label{fig1}}
\end{figure}

The measured cross section $d\sigma/d\Omega$, 
the proton analyzing power $A_y$,  
the $^3\rm He$ analyzing power $A_{0y}$, and the spin correlation
coefficient $C_{y,y}$ are shown in Fig.~\ref{fig1} 
as a function of the center-of-mass (c.m.) scattering angle $\theta_{\rm c.m.}$
together with the theoretical calculations.
The observables for the $p$-$^3\rm He$ elastic scattering were calculated from the solutions 
of exact AGS equations as given in Refs.~\cite{deltuva_prc87, fonseca_FBS2017} using 
a number of  $NN$ potentials:
the Argonne $v_{18}$ (AV18)~\cite{AV18}, 
the CD Bonn~\cite{cdb},  and the INOY04~\cite{INOY04}.
The calculations based on two semilocal momentum space regularized chiral $NN$ potentials
of the fifth order ($\rm N^4LO$)
with the cutoff parameters $\Lambda = 400~{\rm MeV}/c$ (SMS400) 
and  $\Lambda = 500~{\rm MeV}/c$ (SMS500)~\cite{SMS} are also presented.
In addition,
to test the importance of $3N$ and $4N$ forces in the $p$-$^3\rm He$ elastic scattering, 
the calculations based on 
the CD Bonn$+\Delta$ model~\cite{CDBD}, which allows an excitation of a nucleon to
a $\Delta$ isobar and thereby yields effective $3N$Fs and $4N$Fs, 
are presented.

The AGS equations for $4N$ transition operators are solved in the momentum-space partial
wave representation \cite{deltuva_prc87} including $NN$ waves with total angular momentum 
below 4. 
Since the rigorous treatment of the Coulomb force requires 
the inclusion of much higher partial waves,
the Coulomb force is omitted in the present study. 
Given relatively high energy, it is expected to be significant at small angles
up to $\theta_{\rm c.m.} \sim 40^\circ$ only. 

As shown in Fig.~\ref{fig1}, 
the calculations with the $NN$ forces underestimate 
$d\sigma/d\Omega$ at the
backward angles 
$\theta_{\rm c.m.} \gtrsim 80^\circ$.
%
%
It is also found that there is 
a large sensitivity of the calculations to the input $NN$ forces at the minimum region.
INOY04,  which is fitted to reproduce the $^3\rm He$ binding energy, 
provides a better description of the data, 
but it still underestimates the data.  
In addition, $\Delta$-isobar contributions in the 
$p$-$^3\rm He$ elastic scattering,
which are estimated by the difference 
between the CD Bonn+$\Delta$ and CD Bonn calculations, 
are clearly seen to the limited angles $\theta_{\rm c.m.} \lesssim 110^\circ$.
In the $Nd$ elastic scattering, the $\Delta$-isobar effects
increase $d\sigma/d\Omega$ to reduce the discrepancy 
from the data over all angles~\cite{CDBD}.

The calculated $A_y$ 
has a relatively small sensitivity to  $NN$ forces, 
and the description of  $A_y$ 
is moderate.
Small but visible effects of the $\Delta$ isobar 
are predicted by CD Bonn+$\Delta$,
which leads to a better agreement with the data
depending on the measured angles.

As for $A_{0y}$, 
the calculations  based on the $NN$ potentials are close to each other.
The $A_{0y}$ data deviate largely from the $NN$ force
calculations at the minimum $\theta_{\rm c.m.} \sim 90^\circ$
as well as the maximum $\theta_{\rm c.m.} \sim 140^\circ$,
which was not seen at lower energies~\cite{viviani_PRL111,deltuva_prc87, fonseca_FBS2017}.
The $\Delta$-isobar effects shift the calculated results slightly
but in the wrong direction
at $\theta_{\rm c.m.} \sim 100 ^\circ$.

For $C_{y,y}$, the angular dependence looks quite different from that at lower energies
\cite{viviani_PRL111,deltuva_prc87, fonseca_FBS2017}, and 
large $\Delta$-isobar effects are predicted at the angles 
$\theta_{\rm c.m.}=100^\circ$--$140^\circ$.
The data at the limited angles have moderate agreements 
to all the calculations with no definite conclusions
for the $\Delta$-isobar effects.

\section{Discussion}

\begin{figure}[htbp] 
\includegraphics[scale=0.5]{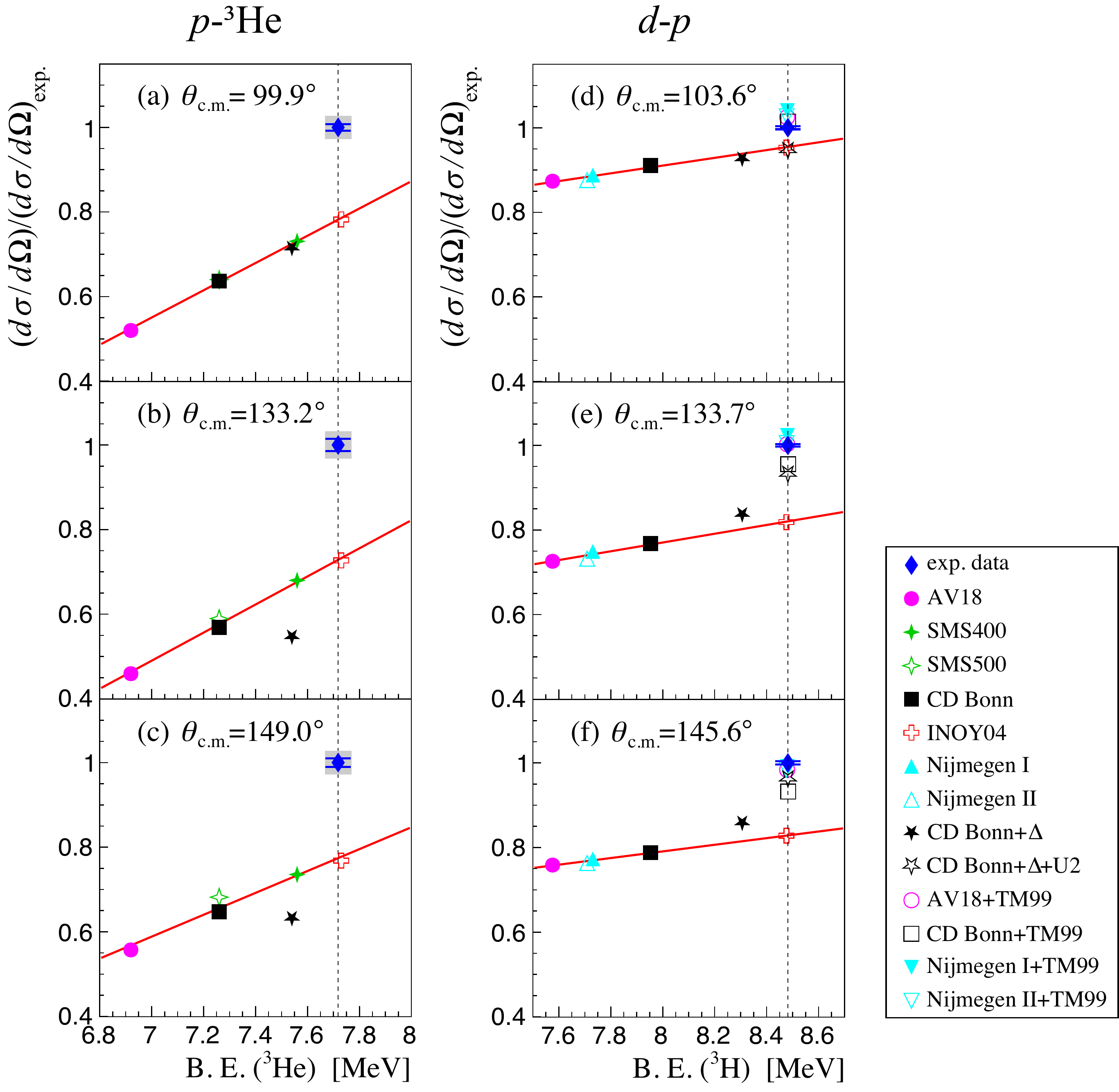}
\caption{(Color online)
Relation between the $^3\rm He$ binding energy and 
the cross section for $p$-$^3\rm He$ elastic scattering at around 65 MeV
in panels (a), (b), and (c);
and that between the $^3\rm H$ binding energy and 
the cross section for $d$-$p$ elastic scattering at 70 MeV/nucleon
in panels (d), (e), and (f).
The result of the cross section for each nuclear potential
is shown as a ratio to the corresponding experimental data.
The dashed vertical straight lines denote the experimental binding energy of the $^3\rm He$ 
in panels (a)--(c) and that of $^3\rm H$ in panels (d)--(f).  
For the scattering data the statistical (systematic) errors are shown with bars (bands).
%
%
%
Correlation lines obtained with the results of the $NN$ potentials
are shown with red lines.
\label{fig3}}
\end{figure}

In Refs.~\cite{deltuva_prc87,fonseca_FBS2017}, 
it was found that  $d\sigma/d\Omega$ 
for the $p$-$^3\rm He$ elastic scattering  calculated
with the AV18, CD Bonn, and INOY04 $NN$ potentials 
at lower energies of 7--35 MeV
scale with the binding energy (B.E.) of $^3$He. 
Inspired by this fact we investigate the scaling relation
between B.E.($^3$He) and the cross section at 65 MeV
in panels (a)--(c) of Fig.~\ref{fig3}. 
In the minimum region of angles 
$\theta_{\rm c.m.}=80^\circ$--$150^\circ$
the calculated  $d\sigma/d\Omega$,
normalized by the corresponding experimental data, 
are plotted as a function of B.E.($^3$He). 
Linear correlations, shown as red straight lines in the figure, 
exist for the calculations based on the $NN$ potentials 
including the two chiral $NN$ potentials, i.e., SMS400 and SMS500.
From the correlation lines one can predict 
the $d\sigma/d\Omega$
corresponding to the $NN$ potential that reproduces the experimental B.E.($^3\rm He$),
which almost coincides with the INOY04 result.
As shown in the figure,
the predictions underestimate the experimental $d\sigma/d\Omega$  
by 20--30\%.
Note that the calculations with CD Bonn+$\Delta$,
that will be discussed later, 
are not included in the fitting of the correlation lines. 

In panels (d)--(f) of Fig.~\ref{fig3}, 
we also demonstrate the scaling relation 
between B.E.($^3\rm H$) and the cross section for the deuteron-proton
($d$-$p$) elastic scattering at 70 MeV/nucleon.
The calculated  $d\sigma/d\Omega$ at 70 MeV/nucleon,
normalized by the experimental data of the $d$-$p$ elastic scattering~\cite{sekiguchi2002},
are plotted as a function of B.E.($^3\rm H$)~\cite{deltuva_2020,Krakow_3H};
the Coulomb interaction is not taken into account.
A similar scaling also exists in the $d$-$p$ elastic scattering
as seen in the $p$-$^3\rm He$ scattering.
The red straight lines are obtained by fitting the calculations with $NN$ potentials: 
AV18, CD Bonn, Nijmegen I, II~\cite{nijm}, and INOY04. 
These lines, as in the case of the $p$-$^3\rm He$ scattering, 
allow us to predict $d\sigma/d\Omega$ with a $NN$ potential that reproduces 
the experimental B.E.($^3\rm H$). 
The panels (d)--(f) of Fig.~\ref{fig3}
show that the predictions 
underestimate the experimental $d\sigma/d\Omega$ by 10--20\%.  
In the figure,
the calculations taking into account 
the Tucson-Melbourne'99 2$\pi$-exchange $3N$F (TM99-$3N$F)~\cite{tm99}
in which the cutoff parameter is fitted to reproduce the experimental B.E.($^3\rm H$)
for each combined $NN$ potential, i.e., AV18, CD Bonn, Nijmegen I and II, 
are presented~\cite{Krakow_3H,sekiguchi2017}.
In addition, 
calculations including an irreducible $3N$ potential 
contribution to CD Bonn+$\Delta$ which reproduces the experimental B.E.($^3\rm H$), 
the model U2 of Ref.~\cite{CDB+Delta+U2}, are also shown.
The calculated results 
provide good agreements with the experimental data, 
indicating strong evidence for the need to include the $3N$Fs.
%
%
The discrepancy between the data and the predictions
with a $NN$ potential that reproduces the experimental B.E.($3N$)
for the $p$--$^3\rm He$ elastic scattering
at the cross section minimum angles 
is similar in size 
or even larger 
than that for the $d$-$p$ elastic scattering 
at a similar incident energy.
It should be interesting to see whether 
the combinations of $2N$ and $3N$ forces, 
that give good descriptions of the $d$-$p$ scattering cross section, 
explain the data for the $p$-$^3\rm He$ scattering. 

An interesting feature found in the above mentioned correlations is that 
dependence of the calculated $d\sigma/d\Omega$ 
on B.E.($3N$) for the $d$-$p$ scattering is smaller than that for the $p$-$^3\rm He$ 
scattering. 
%
It is quantified by the gradients of the correlation lines: $\sim 0.1/\rm MeV$ 
for the $d$-$p$ scattering and $\sim 0.3/\rm MeV$
for the $p$-$^3\rm He$ scattering.
%
There is a speculation that a weaker sensitivity for the $d$-$p$ elastic cross section 
is related to dominance of the total $3N$ spin $S=3/2$ state (quartet state) 
in the nucleon-deuteron elastic scattering~\cite{Koike1986}. 
As for the neutron-deuteron $s$-wave scattering length,
the quartet state is insensitive to the difference 
of $NN$ potentials because of the Pauli principle, which prevents 
two neutrons getting close to each other~\cite{Bahethi1956,Krakow_3H}. 
Therefore it is expected that relatively large dependence seen 
in the $p$-$^3\rm He$ elastic cross sections is a reflection 
of medium- and short-range details 
of the nuclear interactions including $3N$Fs.

In the following, we discuss the $\Delta$-isobar effects.
As shown in panels (a)--(c) of Fig.~\ref{fig3}, 
the calculations of the CD Bonn+$\Delta$ 
model for the $p$-$^3\rm He$ elastic scattering 
are off from the correlation lines at backward angles and move in a direction opposite to the experimental data. 
Thus, the $\Delta$-isobar effects do not improve the agreement with the data. 
Meanwhile,  as shown in panels (d)--(f) of Fig.~\ref{fig3}, 
the calculations of the CD Bonn+$\Delta$ model 
for the $Nd$ elastic scattering are 
off from the correlation lines, but the $\Delta$-isobar effects provide a better agreement with the data. 

\begin{figure}[htbp] 
\includegraphics[scale=0.4]{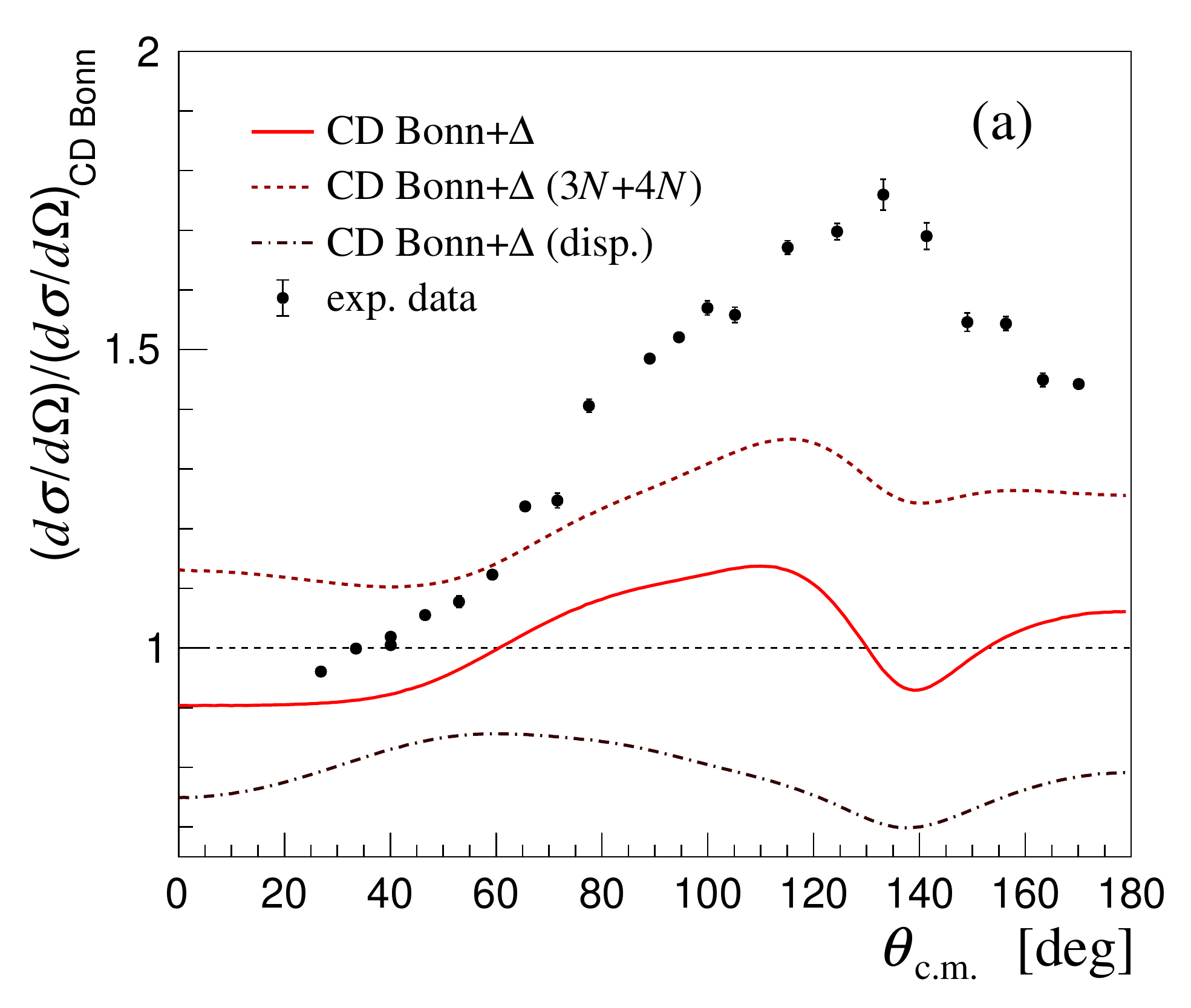}
\includegraphics[scale=0.4]{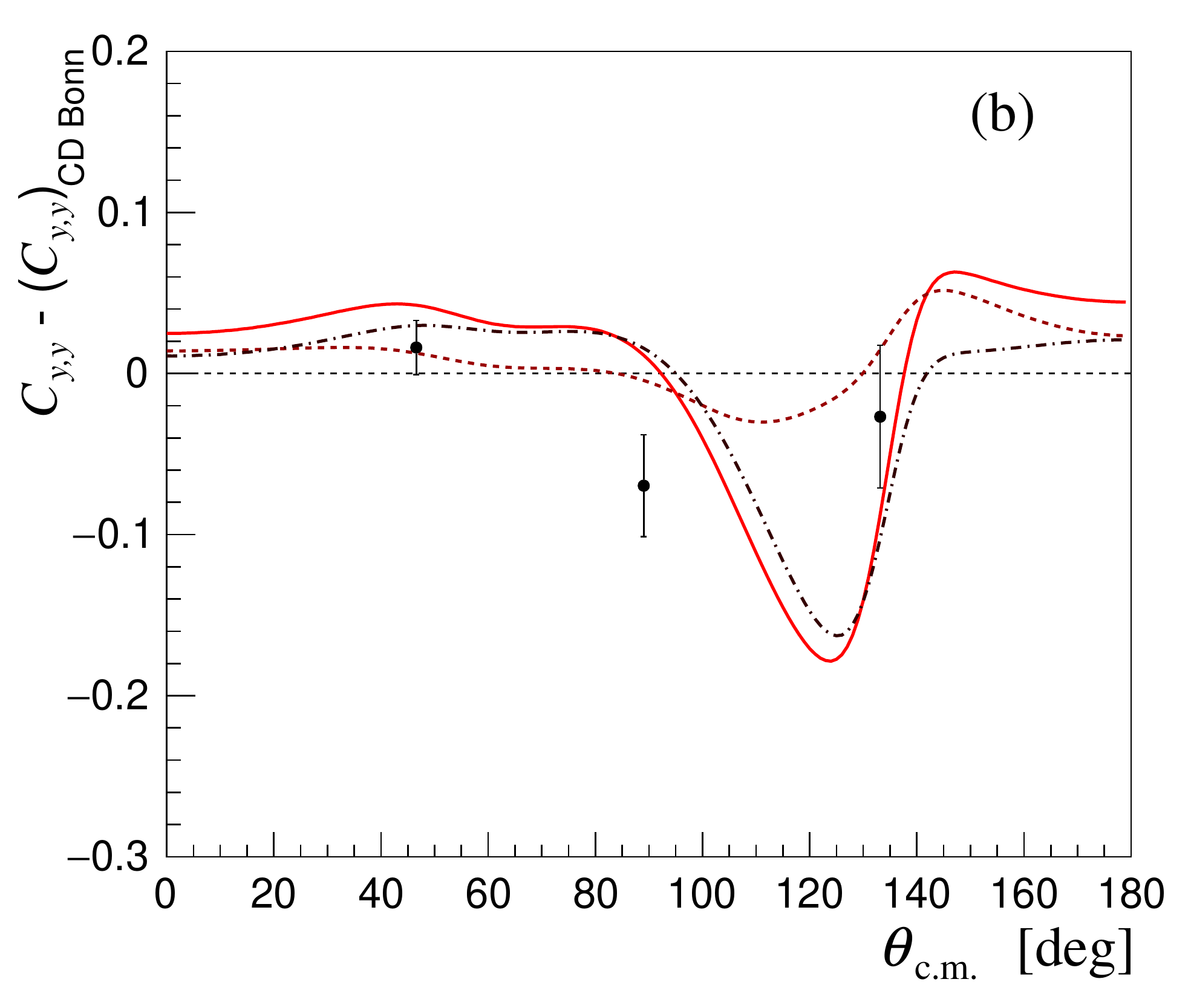}
\caption{(Color online)
Effects of the $2N$ dispersion (dash-dotted lines), those of $3N$- and $4N$-forces 
(dotted lines), and the total $\Delta$-isobar effects (solid lines) 
in the $p$-$^3\rm He$ elastic scattering at 65 MeV
are shown as a function of the c.m. scattering angle 
for the cross section in panel (a) and the spin correlation coefficient $C_{y,y}$  
in panel (b).
For the cross section, 
the result of each contribution is shown as a ratio to the calculation based on the CD Bonn potential. 
For the spin correlation coefficient $C_{y,y}$, 
the differences from the calculation of the CD Bonn potential are presented.
\label{figx}}
\end{figure}

It is known that effective $3N$ and $4N$ forces 
due to the excitation of a nucleon to a $\Delta$ isobar 
are often partially canceled by $\Delta$-isobar effects of $2N$ nature, 
the so-called $2N$ dispersion~\cite{nemoto98, CDBD, CDBD2,Nemoto1999}. 
To study $\Delta$-isobar effects more in detail, 
the effects of the $2N$ dispersion, and those of $3N$ and $4N$ forces, 
are singled out separately as in Ref.~\cite{Deltuva_PRC76,CDBD2}. 
The results for the cross sections are shown in Fig.~\ref{figx}(a)
as a ratio to the calculation based on the CD Bonn potential. 
At the minimum angles, large contributions of the $\Delta$-generated $3N$ and $4N$ forces 
increase the cross section values. 
However, together with this, there is a strong dispersive $\Delta$-isobar effect,
which is opposite to the $3N$F and $4N$F effects. 
As a result, the net effects of the $\Delta$ isobar are small, and their effects 
are even reversed at $\theta_{\rm c.m.}\approx140^\circ$.
In the $Nd$ elastic scattering the dispersive $\Delta$-isobar effect is smaller 
than that of the $\Delta$-generated $3N$Fs, and then the net contributions 
of the $\Delta$ isobar increase the cross section~\cite{nemoto98,Nemoto1999}. 
Since the calculated $3N$ binding energy with CD Bonn+$\Delta$ 
is still smaller than the experimental value by about $0.2~\rm MeV$~\cite{CDBD,CDBD2}, 
further attractive effects attributed to the irreducible $3N$Fs in $NNN$-$NN$$\Delta$ 
model space~\cite{CDB+Delta+U2} should be considered.
It will be interesting to study in the future how such attractive contributions 
affect the cross sections for the $p$-$^3\rm He$ scattering.

Regarding the spin observables,
large $\Delta$-isobar effects are predicted for the spin correlation 
coefficient $C_{y,y}$ (see Fig.~\ref{fig1}).
Interestingly, Fig.~\ref{figx}(b) shows that the predicted $\Delta$-isobar 
effects are mainly due to the $2N$ dispersion. 
Experimental data of the spin correlation coefficient $C_{y,y}$ 
in a wide angular range are needed for a detailed discussion of the $\Delta$-isobar effects. 

\section{Summary and conclusion}

We have reported the precise data set for $p$-$^3\rm He$ elastic scattering
at intermediate energies:
$d\sigma/d\Omega$ and the proton analyzing power $A_y$ 
taken at 65 MeV in the angular regime $\theta_{\rm c.m.} = 26.9^\circ$--$170.1^\circ$;
the $^3\rm He$ analyzing power $A_{0y}$ at $70~\rm MeV$ in the angular regime $\theta_{\rm c.m.} = 46.6^\circ$--$141.4^\circ$, 
and the spin correlation coefficient $C_{y,y}$  at  $65~\rm MeV$ for
$\theta_{\rm c.m.} = 46.6^\circ$, $89.0^\circ$, and $133.2^\circ$.

For the cross section the statistical error is better than $\pm 2\%$ 
and the systematic uncertainties are estimated to be 3\%.  
The absolute values of the cross section were deduced by 
normalizing the data to the $p$-$p$ scattering cross section given 
by the phase-shift analysis program SAID.
For the proton and $^3\rm He$ analyzing powers the statistical uncertainties are less than 0.02.
They are 0.03--0.06 for the spin correlation coefficient $C_{y,y}$.
The systematic uncertainties for all the measured spin observables
do not exceed the statistical ones.

The data are compared with rigorous $4N$-scattering calculations 
based on various realistic $NN$ nuclear potentials without the Coulomb force.
Clear discrepancies have been found for some of the measured observables,
especially in the angular regime around the $d\sigma/d\Omega$ minimum.
Linear correlations exist between the calculations of the $^3\rm He$ 
binding energy and those of $d\sigma/d\Omega$, 
which enables us to evaluate $d\sigma/d\Omega$ with a $NN$ potential 
that reproduces B.E.($^3\rm He$). 
Predicted values of $d\sigma/d\Omega$ 
in the minimum region clearly underestimate the data. 
A similar tendency is obtained in the $Nd$ elastic scattering,
where discrepancies are largely resolved by incorporating $3N$Fs.
The $NN$ potential dependence for the cross section in the $p$-$^3\rm He$ 
scattering is found to be larger than that in the $d$-$p$ elastic scattering, 
which could allow us to anticipate a wealth of information 
on the nuclear interactions from further 
investigation of the $p$-$^3\rm He$ scattering at these energies.

The $\Delta$-isobar effects in the $p$-$^3\rm He$ 
observables are estimated by the $NN$+$N\Delta$ 
coupled-channel approach. 
They do not always remedy the difference between the data and 
the calculations based on the $NN$ potentials.
In the case of $d\sigma/d\Omega$, 
large contributions of the effective $3N$ and $4N$ forces
are largely canceled by the dispersive $\Delta$-isobar effect, that
leads to a rather small total $\Delta$-isobar effect.
The results are in contrast to those in the $d$-$p$ scattering, 
where the cancellation is less pronounced.
Since this approach still misses the $3N$ binding energies, 
its extensions,
e.g., the irreducible 3$N$ potential combined with the $NN$+$\Delta$ model~\cite{CDB+Delta+U2},
are needed. 
Together with this, 
large dispersive $\Delta$-isobar effects predicted 
in the spin correlation coefficient $C_{y,y}$ 
should be investigated experimentally in the future. 

From these obtained results we conclude that $p$-$^3\rm He$ 
elastic scattering at intermediate energies is an excellent tool 
to explore the nuclear interactions including $3N$Fs that could not be accessible in $3N$ scattering. 
Recent study of the $\chi$EFT nuclear potentials intends to use the $d$-$p$ scattering 
data at intermediate energies to derive the higher-order $3N$Fs~\cite{Epelbaum2020}. 
It would be interesting to see how the predictions with such $3N$Fs 
explain the data for the $p$-$^3\rm He$ elastic scattering, 
which will enable us to perform detailed discussions of the effects of $3N$Fs including the $T = 3/2$ isospin channels.

\begin{acknowledgments}
We acknowledge the outstanding work of the accelerator groups
of CYRIC and RCNP for delivering excellent proton beams,
and the RANS team for providing high quality neutron beams.
We thank 
T. Uesaka, 
W. Kim, 
S. Shimoura,  
S. Ota, 
Y. Shimizu, 
T. Kobayashi, 
N. Chiga, 
M. Ohgi,  
M. Fujita, 
T. Averett, 
and S. Yanagawa
for their strong support of the experiments.
S. N. acknowledges support by GP-PU at Tohoku University.
This work was supported finally in part by JSPS KAKENHI Grants
No.JP25105502, No.JP16H02171,  and No.JP18H05404.
%
\end{acknowledgments}

\newpage

\end{document}